\documentstyle{article}

\begin{document}

\newcommand{\r}{{\bf r}}
\newcommand{\half}{\frac{1}{2}}
\newcommand{\gs}{$g_{\rm A}^{(0)}$}
\newcommand{\gt}{$g_{\rm A}^{(3)}$}
\newcommand{\go}{$g_{\rm A}^{(8)}$}
\newcommand{\Gs}{g_{\rm A}^{(0)}}
\newcommand{\Gt}{g_{\rm A}^{(3)}}
\newcommand{\Go}{g_{\rm A}^{(8)}}
\newcommand{\Nc}{$N_{\rm c}$}
\newcommand{\Ne}{N_{\rm c}}
\newcommand{\C}{\cite}
\newcommand{\nc}{\nocite}
\newcommand{\vb}{\vspace{-7 pt}}
\newcommand{\cl}{\centerline}
\newcommand{\dirac}{\!\!\!\not{\!\partial}}
\newcommand{\dir}{\!\!\!\!\!\!\not{\partial}}
\newcommand{\Lma}{$\Lambda_{1}/M$}
\newcommand{\Lmb}{$\Lambda_{2}/M$}
\newcommand{\Eval}{$E_{\rm valence}$}
\newcommand{\Econ}{$E_{\rm cont.}$}
\newcommand{\Etot}{$E$}
\newcommand{\sval}{$\sigma_{\rm valence}$}
\newcommand{\scon}{$\sigma_{\rm cont.}$}
\newcommand{\stot}{$\sigma$}
\newcommand{\kmax}{$k_{\rm max}$}
\newcommand{\ms}{$m_{\rm s}$}
\newcommand{\rf}{$r_{f}\, M$}
\newcommand{\rg}{$r_{g}\, M$}
\newcommand{\beq}{\begin{equation}}
\newcommand{\eeq}{\end{equation}}
\newcommand{\bea}{\begin{eqnarray}}
\newcommand{\eea}{\end{eqnarray}}

\thispagestyle{empty}
\setcounter{page}{0}
\begin{flushright}
{}~~~~~\\
RUB-TPII-54/93 \\
9 August, 1993
\end{flushright}

\vspace{2.5 cm}

\begin{center}
{\large {\bf ROTATIONAL CORRECTIONS TO AXIAL CURRENTS \\
IN   SEMIBOSONIZED SU(3) NAMBU--JONA}-{\bf LASINIO MODEL} }
\end{center}
\vspace{0.37truein}
\centerline{\footnotesize
A. BLOTZ,
M. PRASZA{\L}OWICZ\footnote{
Alexander von Humboldt Fellow, on leave of absence from the
Institute of Physics, Jagellonian
University, ul. Reymonta 4, 30-059 Krak{\'o}w, Poland}
 and K. GOEKE}
\vspace{0.015truein}
\centerline{\footnotesize\it Institute for Theor. Physics II, Ruhr-University}
\centerline{\footnotesize\it  44780 Bochum, Germany}

\vspace{3 cm}

\noindent ABSTRACT

\vspace{0.5 cm}

\noindent We examine $O(1/\Ne)$ rotational corrections to axial couplings
\gt~ and \go~ in the framework of the semibosonized SU(3)
Nambu--Jona-Lasinio model.
A novelty is due to the observation that the rotational (cranking)
velocity and the rotation matrix itself  do not commute within
semiclassical quantization scheme. If time ordering in the quark loop
is maintained new contributions, which have no analogue in the
Skyrme model, appear. They substantially improve
numerical results for the axial couplings which without the present
corrections are badly underestimated.

\newpage

\section{Introduction}

There are now three different measurements of the spin asymmetry in
polarized lepton--nucleon deep inelastic scattering
\C{ga:EMC}\nc{ga:SMC}--\C{ga:E142}. Although the
original experimental papers seemed to show some discrepancies between
SLAC and CERN experiments, the recent work of Ellis and Karliner
\C{ga:ElKarl}
reconciles the three
experiments within one standard deviation. The message of this work is
that whereas the Bjorken sum rule \C{ga:BJ}
is in agreement with the data, the
Ellis--Jaffe sum rule \C{ga:EJ} is violated. Their results read:
\beq
\Gs=0.24\pm 0.09,~~~~\Go=0.35\pm 0.04~~{\rm and}~~\Gt=1.25.
\label{eq:g083}
\eeq

What was at first called a spin crisis has been subsequently understood
as a crisis of the nonrelativistic quark model \C{ga:JM}.
Although the model was quite successful in reproducing \gt~
(the well known value of (\Nc+2)/3) it failed
badly for \gs, predicting -- in our normalization --
\gs$=\sqrt{3}$ \go~ (Ellis--Jaffe sum rule result).
Since the quarks in the nucleon are by
no means free, it soon became clear that the total spin, even if one
neglects gluons and anomaly, has also an orbital component which
screens the nucleon matrix element of the singlet axial current.

In an immediate response to the first EMC result Brodsky, Ellis and
Karliner published a paper \C{ga:BEK}
in which it was shown that the Skyrme model
gives zero for \gs~ unless some extra terms responsible for the
mixing with $\eta^{\prime}$ were added. Although it was in a sense a
welcome result its actual significance was very much weakened by the
fact that the model underestimated \gt~ by a factor of 2
\C{skm:ANW}, (see however \C{skm:PSW}).

In the present paper we report on the calculation of the three axial
couplings: \gs, \gt~ and \go~ in the   semibosonized
Nambu--Jona-Lasinio model (NJL or chiral quark model).
In this model \Nc ~valence quarks are bound
in a self-consistent solitonic field
(see Ref.\C{cqm:Bloc} and references therein). The model interpolates between
the quark model (small soliton size), where the valence quarks are
almost unbound and the Skyrme model, where for large soliton sizes the
valence level disappears in the Dirac sea.
Similarly to the Skyrme model, also in the
NJL model \gt~ was badly underestimated
\C{ga:MeissGo,mp:nucleon}.
The recent result of Refs.\C{ga:WaWa,ga:AllStars} where the
$O$(1/\Nc) correction  to \gt~
was calculated in the SU(2) version of the model seems
to solve this long lasting
problem.

 The new result of the present work consists of a calculation of
$O$(1/\Nc) contributions to \gt~ and \go~ in the SU(3) NJL model.
In fact one expects large $O(1)$ corrections to the leading $O(\Ne)$ term of
for \gt~ and \go; indeed the SU(2) constituent quark model predicts
\gt=(\Nc+2)/3.
In the present model these
corrections are of two types: {\it antisymmetric} ones which have no
counterpart in the Skyrme model since they crucially depend on the time
ordering within the quark loop, and the {\it symmetric} ones which can
be attributed to the Wess-Zumino term. Altogether with \gs, which by
itself gets a contribution only at the $O$(1/\Nc) level
\C{ga:BPG}, the full
results for the  three axial couplings are in relatively good agreement
with experimental data.

\section{Axial Current in the NJL model}

Our  starting point is a semibosonized Euclidean action of
the NJL model with scalar degrees of freedom kept constant \C{cqm:Bloc}:
\beq
S_{\rm eff}[U]=-{\rm Sp} \log
\left\{-i~~\dirac \, +\, m\, +\, M \; U^{\gamma_{5}}\right\},
\label{eq:Seff}
\eeq
where $\gamma$ matrices are antihermitean,
the SU(3) matrix $U^{\gamma_{5}}$ describes chiral fields $\pi$-K-$\eta$,
$M$ is the constituent quark mass
and $m$ denotes  the current quark matrix, which we will put equal
to 0. The influence of the finite quark masses will be investigated
elsewhere \C{ga:BPG2}.  Eq.(\ref{eq:Seff}) has
to be  regularized.
 Following Ref.\C{cqm:Bloc}, we will use the two step cut-off function, with 2
parameters $\Lambda_{1,2}$ to regularize the real part of the Euclidean
action by means of the Schwinger proper-time procedure
The imaginary
part, which is finite to start with,
remains unregularized. This has been at length discussed in
Refs.\C{cqm:Bloc,inst:nucleon}.
 From the gradient expansion of Eq.(\ref{eq:Seff}) in the meson sector
 one fixes  $\Lambda_{1,2}=\Lambda_{1,2}(M)$.
 Therefore the model is very economical: there is in fact only one
explicit parameter, namely $M$.

The model has solitonic solutions
(see Ref.\C{cqm:Bloc} and references therein)
 which  have been successfully interpreted as baryons.
The existence of the soliton is guaranteed due to the interplay between
the valence level and the energy of the continuum states. All
quantities have therefore typically two contributions which we
subsequently call {\it valence} and {\it sea} respectively. The {\it
sea} contributions have usually their counterparts in the Skyrme model.

The splittings within the
SU(3) multiplets of baryons as well as the isospin splittings have been
shown to be in very good agreement with experiment \C{cqm:Bloc,mp:iso}.
To this end one uses the hedgehog Ansatz and performs collective quantization
in terms of the rotation matrix $A(t)$:
$ U(x) = A (t) U_{0}(\vec{x})\, A^{\dagger}(t) $,
where
$ A^{\dagger}\; dA/dt = i/2\;  \sum_{a=1}^{8}
\lambda_{a} \Omega_{a}$,
\beq
U_{0} = {\rm diag}( \overline{U}_{0},1) \label{eq:Ans}
\eeq
and $ \overline{U}_{0}=
\cos P(r) + i \vec{n}  \vec{\tau} \sin P(r) $ with the profile function
satisfying the boundary conditions:
$P(0)=\pi$ and $P(\infty)=0$. The collective SU(3) octet wave functions are
given in  terms of Wigner matrices:
$ D_{a b}(A) = 1/2\;{\rm Tr}(A^{\dagger} \lambda_{a} A \lambda_{b}) $,
where left index goes over all 8 flavor states and the right index
 $b=(Y^{\rm R},T^{\rm R},T_3^{\rm R})$
is confined to the states with $Y^{\rm R}=1$ \C{cqm:Bloc,skm:G}.
The corresponding right isospin
has interpretation of spin ($T_3^{\rm R}=-J_3$).

The space-like collective axial current operator can be
expressed by the following functional  trace in Euclidean space:
\beq
A_{j}^{a}=i\; {\rm Sp} \left\{ \Gamma_{j}^{b} \; D_{a b}
            \frac{1}{\partial_t+H+\frac{i}{2} \Omega_c^{\rm E}
            \lambda_c} \right\}
\label{eq:Jj} \eeq
with $H=-i \gamma_4 (-i \gamma_k \partial_k + M U^{\gamma_5})$ and
$\Gamma_{j}^{b}=\gamma_4 \gamma_j \gamma_5 \lambda_b$. For a singlet axial
current $ \lambda_b$ should be replaced by a unit matrix and
one should also replace
$D_{a b} \rightarrow 1$ in Eq.(\ref{eq:Jj}).

All quantities in the  NJL model are given as  power series in
the cranking velocities
 $\Omega$, where each power of $\Omega$ counts as $1/\Ne$. In
the zeroth order one gets \C{ga:MeissGo}:
\beq
(A_{j}^{a})^{(0)}=
i\; {\rm Sp} \left\{ \Gamma_{j}^{b} \; D_{a b}
            \frac{1}{\partial_t+H} \right\}
\equiv -\half D_{aj}\; A.
\label{eq:Jj0}\eeq
Note that at this level the singlet axial current is equal to 0 in
agreement with the Skyrme model result \C{ga:BEK}.

The next term in $\Omega$ (or in $1/\Ne$ )
is  of great importance.   Expanding in $\Omega$ we get:
\beq
(A_{j}^{a})^{(1)}=\frac{1}{2}\; {\rm Sp} \left\{ \Gamma_{j}^{b} \; D_{a b}
            \frac{1}{\partial_t+H} \Omega^{\rm E}_c
            \lambda_c \frac{1}{\partial_t+H}
       \right\}. \label{eq:Jj1}
\eeq
There are two subtleties connected with Eq.(\ref{eq:Jj1}): first
it should be remembered that because of the collective quantization $\Omega$
is no longer a c-number but rather an operator which does not commute
with $D_{a b}$.
Second, the trace in Eq.(\ref{eq:Jj1}) should
be understood as time-ordered \C{ga:WaWa,ga:AllStars}.

To this end let us consider the Euclidean propagator:
\bea
<x \mid \frac{1}{\partial_t + H} \mid y > & = &
\theta(t_x-t_y)\sum\limits_{E_n>0} \Phi_n(\vec{x})\Phi_n^{\dagger}(\vec{y})
\exp(-E_n(t_x-t_y)) \nonumber \\
 & - &\theta(t_y-t_x)\sum\limits_{E_n<0}
 \Phi_n(\vec{x})\Phi_n^{\dagger}(\vec{y}) \exp(-E_n(t_x-t_y)). \label{eq:prop}
 \eea
If two such propagators are multiplied and time ordering in
Eq.(\ref{eq:Jj1}) is assumed, then the correct expression reads:
\bea
(A_{j}^{a})^{(1)} & = &\frac{1}{2}\;
\sum\limits_{E_m<0 \atop E_n > 0}
\frac{1}{|E_m|+|E_n|}
\left\{
<m \mid \Gamma_{j}^{b} \mid n > < n \mid \lambda_c \mid m >
 D_{a b} \Omega^{\rm E}_c \right.
  \nonumber \\
 & + &
 \left.
 <n \mid \Gamma_{j}^{b} \mid m > < m \mid \lambda_c \mid n >
\Omega^{\rm E}_c D_{a b}
\right\}. \label{eq:OK}
\eea
Taking into account symmetry properties of the matrix elements
and the order of $\Omega$ and $D_{ab}$ in
Eq.(\ref{eq:OK}) we get back in Minkowski space:
\bea
(A_{j}^{a})^{(1)}& = &  [\Omega_c,D_{a b}]\; i
\left\{ \half B\; \epsilon_{jbc} + C\; (f_{jbc}-\epsilon_{jbc})
\right\}  \nonumber \\
& + &\frac{1}{2} \{D_{a b}, \Omega_c\}_+ \;
\left\{ D\; d_{jb^{\prime}c^{\prime}}\;
(\delta_{b^{\prime}b}-\delta_{b^{\prime}8}\delta_{8b})
(\delta_{c^{\prime}c}-\delta_{c^{\prime}8}\delta_{8c})
+ E\; \delta_{jc}\delta_{8b} \right\} .
\label{eq:efd}
\eea
The above structure comes from the general symmetry arguments and from
the specific form of the hedgehog Ansatz (\ref{eq:Ans}).
For the singlet current we get simply:
\beq
(A_{j}^{0})^{(1)} = \sqrt{3} \; \Omega_j E.
\eeq

As we have already mentioned the explicit expressions for constants
$A \ldots F$ have to be regularized if they come from the square of the
Dirac operator in the proper-time scheme or left unregularized otherwise.
With this in mind we get:
\bea
A_{\rm val} & =& ~-~\Ne~ <{\rm val}\mid \lambda_3 \sigma_3\gamma_5
\mid {\rm val}>,
\nonumber \\
A_{\rm sea} & =&
                {N_c \over 2} \sum_{{\rm all}~n}
                <n\mid \lambda_3 \sigma_3\gamma_5 \mid n>
         {\rm sign} (E_n)
         {\cal R}_\Sigma(E_n)
\eea
and
\bea
i \frac{\Ne}{2} \sum\limits_{m\neq n}  {\cal R}_{\cal Q} (E_m,E_n)\;
         { < m \mid \sigma_3 \lambda_a  \mid n > < n \mid \lambda_b
         \mid m > \over \mid E_m - E_n \mid }
          & = & \left\{
\begin{array}{l}
{}~~B~{\rm if}~a=1,b=2, \\
{}~~C~{\rm if}~a=4,b=5,
\end{array} \right. \nonumber \\
-{\Ne} \sum\limits_{m \neq n}  {\cal R}_{\cal M} (E_m,E_n)\;
         { < m \mid \sigma_3 \lambda_a \mid n > < n \mid \lambda_b
         \mid m > \over \mid E_m - E_n \mid }
         & = & \left\{
\begin{array}{l}
{}~~D~{\rm if}~a=b=4, \\
2E~{\rm if}~a=8,b=3
\end{array} \right.
\nonumber \\
 & & \label{eq:BCDE}
\eea
with
\bea
      {\cal R}_{\cal Q} (E_n,E_m) & = &
     \half \bigl({\rm sign} (E_n-\mu_F) - {\rm sign} (E_m-\mu_F) \bigr),
\nonumber \\
      {\cal R}_{\cal M} (E_n,E_m) & = &
     \half \bigl( 1 - {\rm sign} (E_n-\mu_F) {\rm sign} (E_m-\mu_F) \bigr).
           \label{g14}
\eea
The chemical potential $\mu_F$ is chosen in such a way, that it always lies
between  the {\it valence} level and the positive continuum of
states. In this way  $B \ldots E$  contain both
the {\it valence} and the {\it sea} parts simultaneously.
The regularization function:
\beq  {\cal R}_\Sigma(E_n) =
      {1\over \sqrt{\pi} } \int_0^\infty {d\tau \over \sqrt{\tau} }
      e^{-\tau} \phi({\tau\over E_n^2})
\eeq
is given in terms of  the two-step function $\phi({\tau\over E_n^2})$
of Ref.\C{cqm:Bloc} and is identical as in the SU(2) case \C{ga:MeissGo}.

\section{Results for Axial Couplings}

Under the collective quantization scheme angular velocities are promoted to
spin operators
\C{cqm:Bloc,skm:G}:
$\Omega_a \rightarrow J_a/I_1$ for $a=1,2,3$. For
$a=4\ldots 7$ $\Omega_a \rightarrow J_a/I_2$ where operators $J_a$
fulfill SU(3) algebra, but do not correspond to any symmetry, since the
physical states are constrained to
$Y^{\rm R}=-2/\sqrt{3} J_8=1$.
Here $I_{1,2}=O(\Ne)$ are moments of inertia, and can be found in
Ref.\C{cqm:Bloc}.
We get:
$ [J_a, D_{jb}]=i f_{abc} D_{jc}$.
Left generators: $T_a=-D_{ab} J_b$ correspond to flavor.
The proton axial coupling constants:
$g_{\rm A}^{(\alpha)}\equiv <A_3^{(\alpha)}>$
defined for $J_3=\half$ and $T_3=\half$ read:
\bea
g_{\rm A}^{(a)} & = & -\left(A+\frac{B}{I_1}+\frac{C}{I_2}\right)\; D_{a3}
+\frac{D}{I_2}\; \sum\limits_{m=4}^7 d_{3mm}D_{am}J_m
+\frac{E}{I_1}\;D_{a8} J_{3}~~~{\rm for}~a=3,8; \nonumber \\
 g_{\rm A}^{(0)}   & = &~\sqrt{3}~ \frac{E}{I_1}\; J_3.
\eea
Constants $A\ldots
F$ are calculated numerically for the profile function $P(r)$ which
fulfils self-consistent time-independent equation of motion for the
soliton mass. In our notation they are positive.
Matrix elements of the collective operators can be
evaluated with the help of the SU(3) Clebsch--Gordan coefficients and
for the nucleon  they read:
\bea
<D_{88}>~=\frac{3}{10},~~~~& &~~~~<D_{38}>~=\frac{\sqrt{3}}{15} T_3,
\nonumber \\
<D_{83}>~=-\frac{\sqrt{3}}{15} J_3, & &~~~~<D_{33}>~=-\frac{14}{15}T_3 J_3,
\nonumber \\
<d_{3mn}D_{8m}J_n>~=\frac{\sqrt{3}}{30} J_3, & &
<d_{3mn}D_{3m}J_n>~=\frac{14}{30} T_3 J_3.
\eea

For the purpose of numerical illustration we present in Table 1 results
for $M= 420$~MeV
(this value follows from the overall fit to the hyperon spectra \C{cqm:Bloc})
and, for comparison, also for $M=380$~MeV.
\begin{table}[h]
\begin{center}
\begin{tabular}{|c|c|c|c|c|c|c|} \hline
constans of Eq.(15)       &
 \multicolumn{3}{|c|}{$M=380$~MeV}&\multicolumn{3}{|c|}{$M=420$~MeV} \\
 \cline{2-7}
contribiting to $g_{\rm A}$
          & \gt  & \go  & \gs  & \gt  & \go  & \gs  \\ \hline \hline
$ A $
             &0.63  & 0.15 & --   & 0.59 & 0.15 & --  \\
$D$ and/or $E$
            & 0.28 & 0.14 & 0.42   & 0.26 & 0.13 & 0.37  \\
$B$ and $C$
            & 0.69~(0.43) & 0.17~(0.11) & --
            & 0.73~(0.45) & 0.18~(0.11) & --  \\ \hline
total       & 1.60~(1.34) & 0.46~(0.40) & 0.42
            & 1.58~(1.30) & 0.45~(0.39) & 0.37  \\
\hline
 \end{tabular}
\end{center}
\caption{Various contributions to $g_{\rm A}$, figures in parenthesis
         correspond to the {\it valence} part of constants $B$ and $C$.}
\label{tab}
\end{table}

The above calculation shows that $1/\Ne$ cranking corrections to
$g_{\rm A}$ are not negligible and, in accordance with the
naive expectations
drawn from the quark model, relatively large.
Constants $B$ and $C$ emerge due to the time ordering within the quark
loop. Although their {\it sea} contributions are not small, they have
no corresponding counterparts in the Skyrme model.
These new conributions split almost equally between the purely SU(2)
correction ($B/I_1$) and the purely SU(3) part ($C/I_2$).
Let us stress that we have calculated constants
$B$ and $C$ from the unregularized expressions of Eq.(\ref{eq:BCDE}).
We postpone the discussion of possible
regularization schemes of $B$ and $C$
which would be compatible with the time ordering of the path integral
to the forthcoming paper \C{ga:BPG2}. Here we content ourselves
with simply removing the  {\it sea } part from $B$ and $C$ and
quote in parenthesis in Table 1 the  contributions
corresponding only to the {\it valence} parts of $B$ and $C$.
Constants $D$ and $E$ need not to be
regularized; in the Skyrme model they follow from
the Wess-Zumino  term \C{skm:PSW},
(however $E=0$ for the hedgehog Ansatz).

These are, however,
 by no means the only $1/\Ne$ corrections one can think of. First of
all pion loop corrections have to be in principle taken into account.
In the Skyrme model they have been shown to be large and negative
\C{skm:softpi}.
There is also
some dispute in the literature concerning the value of the axial coupling
of the constituent quark \C{cqm:Wein1}, which in the present
formulation of the model
is equal to 1 (see however Refs\C{mp:solitons,mp:ga} and
references therein).
Many different calculations indicate that it might
be actually smaller than 1.
Our results have to be  therefore understood
as the first step towards $1/\Ne$ corrections. It is however
encouraging that we get a positive contribution which
slightly overestimates the values of axial couplings, so that there is
place for negative corrections from  other sources. Another problem
is the departure from the chiral limit. Here
the situation as far as cranking corrections are concerned is rather
straightforward: one has to expand Eq.(\ref{eq:Jj}) not only in terms of
cranking velocity but also in terms of the current mass matrix $m$. Our
preliminary calculations indicate that these corrections
are  small
although not always negligible; this will be a subject of a
separate paper \C{ga:BPG2}.

Let us also note that taking for our Ansatz $U$  the SU(3) matrix
rather than U(3), we have neglected the contribution of $\eta^\prime$.
It was shown in Ref.\C{cqm:Kato} that the influence of the additional
 U(1) factor and the 't~Hooft interaction  on the solionic quantities
 is for all practical purposes  rather small.

\vspace{0.3 cm}

\noindent {\bf Acknowledgements}

\noindent We would like to thank V.Yu. Petrov for discussion
which stimulated this work and P.V. Pobylitsa for discussion and numerous
technical remarks.
We also thank
M.V. Polyakov, M. Wakamatsu, T. Watabe and
Ch. Christov.
The work was partially supported  by
{\it Polish Research Grant} KBN-2.0091.91.01  (M.P.),
{\it Graduiertenstipendium des Landes NRW}
and {\it Ruth und Gerd Massenberg-Stiftung}
(A.B.).

\end{document}